\documentclass[12pt]{article}
\usepackage[top=30truemm,bottom=28truemm,left=25truemm,right=25truemm]{geometry}

\usepackage{indentfirst}
\usepackage[dvipdfmx]{graphicx}
\usepackage{float}
\usepackage{url}
\usepackage{siunitx}
\usepackage{array}
\usepackage{amsmath}
\usepackage{amssymb}
\usepackage{fancybox}
\usepackage{ascmac}
\usepackage{bm}
\usepackage{mathrsfs}
\usepackage{cancel}
\usepackage{physics}
\usepackage{color}
\usepackage{authblk}
\usepackage{appendix}
\usepackage{comment}
\usepackage{mhchem}
\usepackage{caption,latexsym,amsfonts,mathtools,here}
\usepackage{subcaption}
\usepackage{hyperref}

\numberwithin{equation}{section}

\newcommand{\no}{\nonumber}
\newcommand{\h}{\hspace{1mm}}
\newcommand{\hhh}{\hspace{6mm}}

\newcommand{\pa}{\partial}

\begin{document}

%==========================================================
\begin{titlepage}
\begin{flushright}
{\small KOBE-COSMO-23-10}
\end{flushright}

\vspace{50pt}

\begin{center}

{\large{\textbf{Search for  high-frequency gravitational waves with Rydberg atoms}}}

\vspace{25pt}

{Sugumi Kanno$^*$, Jiro Soda$^{\flat, \sharp}$, and Akira Taniguchi$^*$}
\end{center}

\vspace{20pt}

\shortstack[l]
{\hspace{1.2cm}\it $^*${\small Department of Physics, Kyushu University, Fukuoka 819-0395, Japan}\\[4pt]
\it \hspace{1.2cm}$^\flat${\small Department of Physics, Kobe University, Kobe 657-8501, Japan}\\[4pt]
\it \hspace{1.2cm}$^\sharp${\small International Center for Quantum-field Measurement Systems for Studies}\\[4pt]
\it {\hspace{1.15cm}of the Universe and  Particles (QUP), KEK, Tsukuba 305-0801, Japan}}

\vspace{28pt}

%==========================================================
\begin{abstract}
We propose high-frequency gravitational wave (GW) detectors with Rydberg atoms. Rydberg atoms are ultra-sensitive detectors of electric fields. By setting up a constant magnetic field, a weak electric field is generated upon the arrival of GWs. The weak electric field signal is then detected by an electromagnetically induced transparency (EIT) in the system of the Rydberg atoms. Recently, the minimum detectable electric field with the Rydberg atoms is further improved by employing superheterodyne detection method. Hence, even the weak signal generated by GWs turns out to be detectable. We calculate the amplitude of Rabi frequency of the Rydberg atoms induced by the GWs and show that the sensitivity of the Rydberg atoms becomes maximum when the size of the Rydberg atoms is close to the wavelength of GWs. We evaluate the minimum detectable  amplitude of GWs with Rubidium Rydberg atoms and find that the detector can probe GWs with a frequency $f=4.2~\si{GHz}$ and an amplitude around $ 10^{-20}$. %We also find this detector can be used for detecting gravitational waves with frequencies in the range $f=0.3 \sim 16~\si{GHz}$.
\end{abstract}
%==========================================================
\end{titlepage}

\tableofcontents

%==========================================================
%%%%%%%%%%%%%%%%%%%%%%%%%%%%%%%%%%%%%%%%%%%%%%%%%%%%
\section{Introduction}
The discovery of gravitational waves (GWs) from a merging black hole binary by LIGO-Virgo Collaboration~\cite{LIGOScientific:2016aoc}  triggered a new field of science, so-called GW astronomy.  
As the history of astronomy tells us, it would be crucial to expand the range of observable GW frequency for making further discoveries. Currently, LIGO/Virgo/KAGRA are sensitive to GWs with frequencies only in the range of $10\sim10^2$ Hz~\cite{KAGRA:2013rdx}. The future space-based GW observatories such as LISA~\cite{LISA:2017pwj} and DECIGO~\cite{Seto:2001qf,Yagi:2011wg} will cover the low-frequency band $10^{-4}\sim 10^{-1}$ Hz. Moreover,  Pulsar timing arrays are expected to operate in the lower-frequency band $10^{-9}\sim10^{-7}$ Hz. Recently, it has been reported that the pulsar timing arrays have successfully observed GWs~\cite{NANOGrav:2023gor,NANOGrav:2023hde,Reardon:2023gzh,Reardon:2023zen,Zic:2023gta,Antoniadis:2023ott,Antoniadis:2023lym,Antoniadis:2023xlr,Xu:2023wog}. Thus, GWs in the frequency range from $10^{-9}$ Hz to $10^2$ Hz have been well explored.

On the other hand, observations of GWs at frequencies higher than $10$ kHz have not been well developed.  
This is partially due to a tacit assumption that GWs with a frequency higher than 10 kHz have no relevance to physics. However, the high-frequency GWs associated with astrophysical phenomena could exist~\cite{Saito:2021sgq,Ito:2023fcr}. Moreover, primordial black holes (PBHs) lighter than solar mass can produce  GWs with frequencies higher than 10 kHz. We can also expect high-frequency GWs from inflation~\cite{Ito:2016aai,Ito:2020neq}. The frequency of primordial GWs generated by reheating is typically in the range from MHz to GHz~\cite{Ito:2020neq}. High-frequency GWs could be generated by cosmological events even after inflation such as the cosmological phase transitions in the early universe~\cite{Caprini:2018mtu}. In addition, we consider GWs from extra dimensions as the possible candidates for GW sources that could lead to verifying theories beyond the Standard Model. According to~\cite{Clarkson:2006pq}, GWs generated by pointlike bodies orbiting a braneworld black hole can produce observable high-frequency GWs. Hence, observations of high-frequency GW will provide information beyond the Standard Model of particle physics (see a review article~\cite{Aggarwal:2020olq}).
 
Another reason why GWs with frequencies higher than $10$ kHz have not been explored is the presence of obstacles in their detection. Indeed, it is known that the sensitivity of current GW detectors becomes worse in the high-frequency range~\cite{Akutsu:2008qv}.
Hence, it is necessary to come up with new schemes for the detection of GWs at frequencies above $10$ kHz. A hint can be obtained by focusing on the similarities between axions and GWs. That is, if axions can be detected, it opens the possibility of detecting GWs as well. Based on this idea, the use of axion detection with magnons provided constraints on high-frequency GWs~\cite{Ito:2019wcb,Ito:2020wxi}. This approach can be applied to the interaction between axions and various other excitations, such as axion-photon conversion \cite{Ejlli:2019bqj,Domcke:2022rgu}. 
Recently, an idea for the detection of axion by using Rydberg atoms was proposed in~\cite{Engelhardt:2023qjf}. Following the strategy mentioned above, we focus on the similarities between axions and GWs, and then see if  
we can exploit the Rydberg atoms for detecting high-frequency GWs. 

Since Rydberg atoms can exhibit very large electric dipole moments~\cite{gallagher1994rydberg}, they are widely used to measure electric fields as quantum sensors~\cite{degen2017quantum}.  Indeed, it is possible to measure microwave electric fields over a wide range of frequencies from kHz~\cite{jau2020vapor} to THz~\cite{wade2017real,downes2020full}. The Rydberg atoms can be used as GW detectors because GWs induce a weak electric field when propagating in a homogeneous constant magnetic field.
Recently, highly sensitive detection 
utilizing electromagnetically induced transparency (EIT) with an atomic superheterodyne receiver has been developed~\cite{jing2020atomic}. 
Therefore, it would be worth investigating the possibility of detecting high-frequency GWs with the Rydberg atoms.\footnote{See the literature \cite{fischer1994transition} that considers Rydberg atoms to detect 100Hz GWs as an earlier work.}
In particular, the detector with Rydberg atoms 
has a potential for utilizing quantum entanglement to realize Heisenberg scaling.

The paper is organized as follows. In section 2, we explain our strategy for detecting GWs based on the fact that the electric field can be induced by GWs in the presence of a magnetic field.  In section 3, we start with basics of Rabi oscillations. We derive the relation between the susceptibility and density matrix of Rydberg atoms. Then, by solving the master equation, we explain a mechanism of EIT. In section 4, we review the basic idea of Rydberg atom superheterodyne receiver and describe how to use it as a GW detector. Moreover, we review the full-order Fermi normal coordinates for plane GWs. In section 5, 
we deduce the density matrix by solving the Lindblad master equation for the system of Rydberg atom. Then, we evaluate the sensitivity of GW detectors with Rydberg atom. We also discuss the possibility of Heisenberg scaling to enhance the sensitivity.
The final section is devoted to conclusion. In appendix, the details of calculations for the averaged amplitude of GW are presented. We work in natural unit: $\hbar=c=\varepsilon_0=\mu_0=1$.

%==========================================================
%==========================================================
\section{A strategy for detecting GWs}
In this section, we propose a method for detecting high-frequency GWs.
In the presence of a uniform magnetic field, GWs induce
 electric fields. The idea is to detect the electric fields by a heterodyne detection method with Rydberg atoms~\cite{jing2020atomic}.

We first derive the electric field induced by GWs. Below we work in the laboratory frame. We consider a situation where GWs propagate in a constant magnetic field. The action we consider is the electromagnetic fields expressed by
    \begin{eqnarray}
    \label{action}
    S=\int d^4x\sqrt{-g}\left[-\frac{1}{4}F^{\mu\nu}F_{\mu\nu}\right]\ .
    \end{eqnarray}
We consider the tensor mode perturbation $h_{\mu\nu}$ in the four-dimensional metric:
    \begin{eqnarray}
    \label{metric}
    ds^2=(\eta_{\mu\nu}+h_{\mu\nu})dx^\mu dx^\nu\ ,
    \end{eqnarray}
where $\eta_{\mu\nu}={\rm diag}(-1, 1, 1, 1)$ represents the Minkowski metric. The indices $(\mu, \nu)$ run from 0 to 3, and $(0, 1, 2, 3) = (t, x, y, z)$. Substituting Eq.~(\ref{metric}) into the action (\ref{action}), the action up to the first order in $h_{\mu\nu}$ is given by
    \begin{eqnarray}
    \label{maxwell}
    S&=&\int d^4x \left[-\frac{1}{4}F^{\mu\nu}F_{\mu\nu}-\frac{1}{4}\left(\frac{1}{2}hF^{\mu\nu}F_{\mu\nu}+h^\nu{}_{\alpha}F^{\alpha\mu}F_{\mu\nu}-h^\mu{}_{\alpha}F^{\alpha\nu}F_{\mu\nu}\right)\right]\ ,\label{action_h}
    \end{eqnarray}
where $F^{\mu\nu}=\pa^{\mu}A^\nu-\pa^\nu A^\mu$ is the field strength of a gauge field $A^\mu$. The variation of the action (\ref{action_h}) with respect to the gauge field $A^{\mu}$ gives the Maxwelll equation,
    \begin{eqnarray}
    \label{maxwell}
    \pa_\nu F^{\mu\nu}+\pa_\nu\left(\frac{1}{2}hF^{\mu\nu}+h^{\nu}{}_{\alpha}F^{\alpha\mu}-h^\mu{}_{\alpha}F^{\alpha\nu}\right)=0\ .
    \label{Maxwell}
    \end{eqnarray}
If we consider the situation that the electric field $E^i$ is induced by the interaction between the constant magnetic field $B^i$ and GWs, then  $B^i\sim\mathcal{O}(1)$ and $E^i\sim\mathcal{O}(h)$. 
Eq.~(\ref{Maxwell}) at the first order in $h$ becomes
\begin{eqnarray}
    \pa_i \left(E^i-h^0{}_{j}\varepsilon^{jik}B^k\right)=0 \ ,
\end{eqnarray}
where we used the relations $F^{0i}=E^i$ and $F^{ij}=\varepsilon^{ijk}B^k$. 
This can be integrated as
\begin{eqnarray}
    E^i-h^0{}_{j}\varepsilon^{jik}B^k=\varepsilon^{ijk}\pa_jC^k\,,
\end{eqnarray}
where $C^k$ is an arbitrary constant function. If we impose the initial condition that no electric field is induced without the magnetic field, then $C^i=0$. 
Thus, we obtain
    \begin{eqnarray}
    \label{inducedE}
    E^i{}=\varepsilon^{ijk}h^0{}_j{B}^{k}\ .
    \end{eqnarray}
We can naively guess that the magnitude of the induced electric fields is of the order of
\begin{eqnarray}
E^i{}=10^{-17}\,\left(\frac{h}{10^{-20}}\right)\left(\frac{B}{10\,{\rm T}}\right)\,{\rm eV}^2 \sim 10^{-13}\,\left(\frac{h}{10^{-20}}\right)\left(\frac{B}{10{\rm T}}\right)\,{\rm V/cm}.
\end{eqnarray}
Our goal is to study the details of the detection method  and  clarify if we can detect this tiny electric fields with Rydberg atoms.
%==========================================================
\section{Measurement of electric field with EIT}
In the previous section, we showed the strategy for detecting gravitational waves by measuring the electric field. In this section, we review the concept of Electromagnetically induced transparency (EIT) in quantum optics that can be used to measure electric field signals~\cite{scully1999quantum}.
%%%%%%%%%%%%%%%%%%%%%%
%%%%%%%%%%%%%%%%%%%%%%%%%%%%%%%%%%%%%
\begin{figure}[h]
        \vspace{-0.5cm}
\begin{minipage}[b]{0.4\linewidth}
        \centering
        \includegraphics[width=8cm]{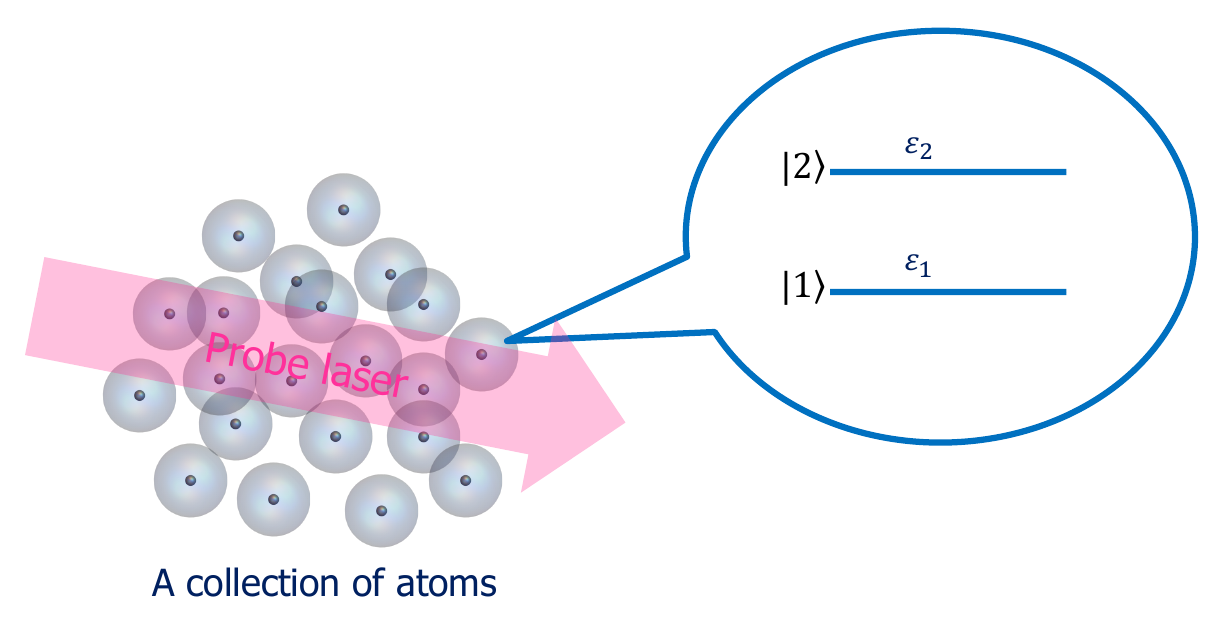}
        \vspace{-2pt}
        \subcaption{\small Sketch of the spectroscopic setup.}
\end{minipage}
  \begin{minipage}[b]{0.7\linewidth}
        \label{fig:2level}
    \centering
    \includegraphics[width=7.5cm]{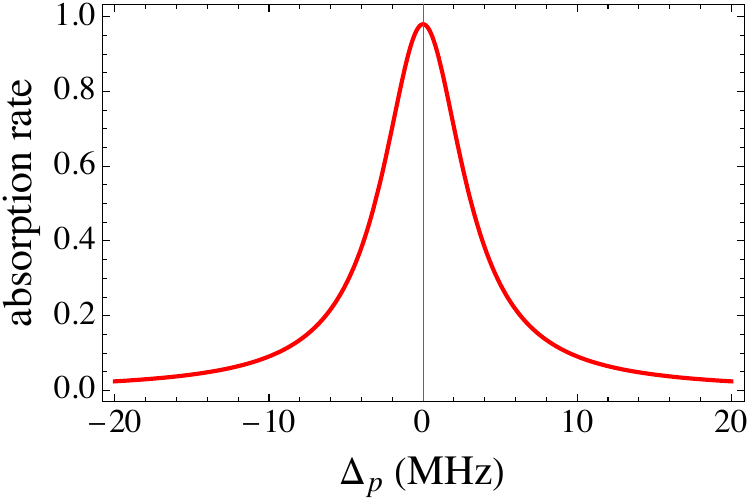}
    \vspace{-4pt}
    \subcaption{The absorption rate of the probe laser.}
  \end{minipage}
    \caption{\small (a) The spectroscopic setup where probe laser is directed at the collection of two-level Rydberg atoms. (b) The absorption rate of the probe laser as a function of the detuning $\Delta_p$. The probe laser is shown to be absorbed by the atoms as $\Delta_p$ goes to zero.}
\vspace{-0.6cm}
\end{figure}
%%%%%%%%%%%%%%%%%%%%%%%%%%%%%%%%%%%%%
\subsection{Interaction of a two-level atom with an electric field}
We first review the interaction system between two-level atoms and the probe laser. Let us consider a situation where the two-level atom with energies $\varepsilon_1$ and $\varepsilon_2$ interacts with an electric field (probe laser) $\bm{E}_p\cos(\omega_p t)$ as shown in Fig.\,1 (a). The total Hamiltonian for this system can be written as 
\begin{eqnarray}
\hat{H}=\hat{H}_0+\hat{H}'(t)\ ,
\end{eqnarray}
where $\hat{H}_0$ represents an unperturbed Hamiltonian and $\hat{H}'(t)$ represents an interaction Hamiltonian. They are written as
\begin{eqnarray}
\hat{H}_0&=&\varepsilon_1\ket{1}\bra{1}+\varepsilon_2\ket{2}\bra{2}\ ,\\[6pt]
\hat{H}'(t)&=&-\hat{\bm{d}}\cdot\bm{E}_p\cos(\omega_p t)\ ,
\end{eqnarray}
where $\hat{\bm{d}}\equiv e\hat{\bm{r}}$ is the electric dipole moment operator and is expressed as
\begin{eqnarray}
\label{d12}
\hat{\bm{d}}=\bm{d}_{\rm 12}\ket{1}\bra{2}+\bm{d}_{\rm 21}\ket{2}\bra{1}\ .
\end{eqnarray}
Here $\bm{d}_{\rm 12}=\bra{1}e\hat{\bm{r}}\ket{2}$ and $\bm{d}_{\rm 21}=\bra{2}e\hat{\bm{r}}\ket{1}$.  
Note that $\bm{d}_{ij} $ represent the electric dipole moment associated with the transition from $\ket{i}$ to  $\ket{j}$.
Thus, we can write the interaction Hamiltonian as follows:
\begin{eqnarray}
\hat{H}'(t)&=&-\hat{\bm{d}} \cdot \bm{E}_p\cos(\omega_p t)\no\\[6pt]
&=&-\frac{1}{2}\left(\bm{d}_{\rm 12}\cdot\bm{E}_p\ket{1}\bra{2}+\bm{d}_{\rm 21}\cdot\bm{E}_p\ket{2}\bra{1}\right)\left(e^{i\omega_p t}+e^{-i\omega_p t}\right)\no\\[6pt]
&=&-\frac{\Omega_p}{2}e^{-i\omega_p t}\ket{2}\bra{1}-\frac{\Omega_p}{2}e^{i\omega_p t}\ket{1}\bra{2} \ ,
\end{eqnarray}
where we used the rotation wave approximation, namely, the terms $e^{-i\omega_p t}\ket{1}\bra{2}$ and $e^{i\omega_p t}\ket{2}\bra{1}$ are ignored. This is because $e^{-i\omega_p t}\ket{1}\bra{2}$ represents a transition in which a photon is absorbed and the energy level drops from $\ket{2}$ to $\ket{1}$, and $e^{i\omega_p t}\ket{2}\bra{1}$ represents a transition in which a photon is emitted and the energy level goes up from $\ket{1}$ to $\ket{2}$, which are physically suppressed.

We note that the transition from $\ket{1}$ to $\ket{2}$ can be realized by absorbing the probe laser with the frequency corresponding to the energy gap of the two levels $\varepsilon_2 -\varepsilon_1$. The probe laser $\bm{E}_p\cos\left(\omega_pt\right)$ can induce the Rabi oscillation with the Rabi frequency  
$\Omega_p\equiv|\bm{d}_{12}\cdot\bm{E}_p|$.
%Here, $\Omega_p\equiv|\bm{d}\cdot \bm{E}|$ is the Rabi frequency which describes the Rabi oscillation between the level $\ket{1}$ and the level $\ket{2}$ induced by the probe laser. 
In this two-level system, the atoms absorb the probe laser in the resonance condition $\Delta_p=0$ as shown in Fig.\,1 (b). %As the result, Rabi oscillation turns out to  be induced between the two-level atoms.
%%%%%%%%%%%%%%%%%%%%%%
\subsection{Susceptibility and density matrix}
In this subsection, we show that the absorption rate of the probe laser and the density operator of the system are related through the complex susceptibility. Specifically, it is known that the absorption rate is proportional to the imaginary part of electric susceptibility which can be related to the density operator of an atom.

First, we show that the absorption rate is proportional to the imaginary part of electric susceptibility. Let us consider the electric field of a plane electromagnetic wave traveling in the $z$-direction
\begin{eqnarray}
\label{electric}
E(z,t)=E_0 \exp(ikz-i\omega t)\,,
\end{eqnarray}
where $E_0$ is the amplitude of the electric field. The complex relative permittivity $\varepsilon(\omega,k)$ is expressed as
\begin{eqnarray}
\label{1+chi}
\varepsilon(\omega,k)=1+\chi(\omega,k)\ ,
\end{eqnarray}
where $\chi(\omega,k)$ ( $|\chi|\ll 1$ ) is the linear susceptibility. We can introduce the refractive index $n(\omega)$ and the extinction coefficient $\kappa(\omega)$ by separating $\sqrt{\varepsilon(k, \omega)}$ into its real and imaginary parts such as
\begin{eqnarray}
\label{n+ik}
\sqrt{\varepsilon(k, \omega)}=\frac{k}{\omega}\equiv n(\omega)+i\kappa(\omega)\ .
\end{eqnarray}
Then Eq.~(\ref{electric}) is written as
\begin{eqnarray}
\label{E}
E(z,t)=E_0\exp\left[i\left(\omega n(\omega)z-\omega t\right)\right]\exp\left[-\omega\kappa(\omega)z\right]\ .
\end{eqnarray}
Note that $\kappa(\omega)$ represents an exponential decay of the electric field, so it is referred to as absorption rate. Next, defining  real and imaginary parts of the susceptibility as {$\chi'$}and $\chi''$, we can write  
\begin{eqnarray}
\label{chi}
\chi \equiv \chi' + i \chi''\ .
\end{eqnarray}
Substituting Eq.~(\ref{chi}) into Eqs.~(\ref{1+chi}) and (\ref{n+ik}), we find
\begin{eqnarray}
\kappa(\omega)=\frac{\chi''}{2}\ .
\label{kappa}
\end{eqnarray}
This shows that the absorption rate is proportional to the imaginary part of susceptibility.
%=========================================================

Next, we derive the relation between the susceptibility and density matrix of an atom. Let us focus on a two-level system  $\ket{1}$ and $\ket{2}$ of an atom. In the semiclassical theory, the polarization $\bm{P}$ at the position of an atom induced by the probe laser $\bm{E}_p\cos\left(\omega_p t\right)$ is expressed as $\bm{P}(t)=N\Tr[\hat{\bm{d}}\rho(t)]$, where $N$ is the atom number density and $\rho(t)$ is the density operator of the atom. Combining this with Eq.~(\ref{d12}), we have
    \begin{eqnarray}
    \bm{P}(t)%=N\Tr[\hat{\bm{d}}\rho(t)]
    =N\bm{d}_{12}\rho_{21}(t)+{\rm~c.c.}\ .
    \label{Pol}
    \end{eqnarray}
If we decompose the polarization $\bm{P}(t)$ into Fourier modes, we have
    \begin{eqnarray}
    \label{PFourier}
    \bm{P}(t)%=\tilde{\bm{P}}(\omega_p)e^{-i\omega_pt}+\tilde{\bm{P}}(-\omega_p)e^{i\omega_pt}
    =\tilde{\bm{P}}(\omega_p)e^{-i\omega_pt}+{\rm c.c.}\ .
    \end{eqnarray}
where $\tilde{\bm{P}}(\omega_p)$ is the Fourier transform of $\bm{P}(t)$. By comparing Eq.~(\ref{Pol}) with Eq.~(\ref{PFourier}), we find
    \begin{eqnarray}
    \label{compNdrho}
    \tilde{\bm{P}}(\omega_p)=N \bm{d}_{12} \rho_{21}(t)e^{i\omega_p t}\ .
    \end{eqnarray}
Similarly, the electric field $\bm{E}(t)=\bm{E}_p\cos\left(\omega_pt\right)$ is decomposed into Fourier mode %with $\pm\omega_p$:
    \begin{eqnarray}
    \label{EFourier}
    \bm{E}(t)=\tilde{\bm{E}}(\omega_p)e^{-i\omega_pt}+{\rm c.c.}=\frac{1}{2}\bm{E}_pe^{-i\omega_pt}+{\rm c.c.}
    \end{eqnarray}
with $\tilde{\bm{E}}(\omega_p)$ being the Fourier transform of $\bm{E}(t)$. Note that we used $\cos\left(\omega_pt\right)=\left(e^{i\omega_pt}+e^{-i\omega_pt}\right)/2$ in the second equality.
%Focusing on the mode of $+\omega_p$ in Eq.~(\ref{PFourier}) and Eq.~(\ref{EFourier}), 
The susceptibility is defined by 
    \begin{eqnarray}
    \label{DIFsus}
    \tilde{\bm{P}}(\omega_p)=\chi(\omega_p)\tilde{\bm{E}}(\omega_p)\,.
    \end{eqnarray}
So if we use Eqs.~(\ref{compNdrho}), (\ref{EFourier}) and (\ref{DIFsus}), the susceptibility is found to be related to the density operator of the atom
    \begin{eqnarray}
    \label{chirho}
    \chi(\omega_p)=\frac{2N|\bm{d}_{12}|^2}{\Omega_p}\rho_{21}(t)e^{i\omega_p t}\ .
    \end{eqnarray}
From the above equations Eq.~(\ref{kappa}) and Eq.~(\ref{chirho}), we can see that the absorption rate can be evaluated from the density operator. Therefore, in the next step, we present a specific system and find its density operator.

%%%%%%%%%%%%%%%%%%
\begin{figure}[t]
\label{fig:3levelsys}
  \begin{minipage}[b]{0.4\linewidth}
    \centering
    \includegraphics[width=6.5cm]{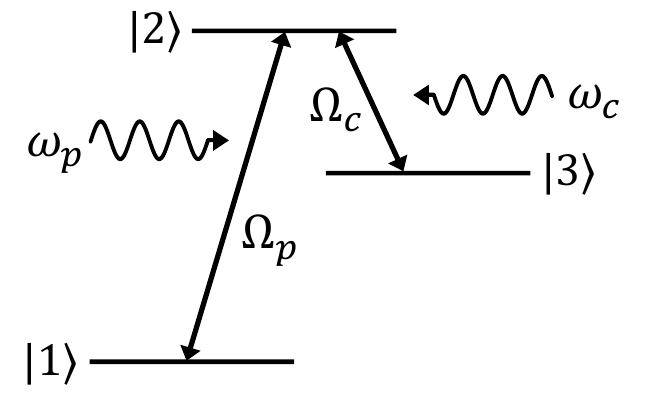}
    \vspace{14pt}
    \subcaption{The energy level of EIT system.}
  \end{minipage}
  \begin{minipage}[b]{0.55\linewidth}
    \centering
    \includegraphics[width=8.5cm]{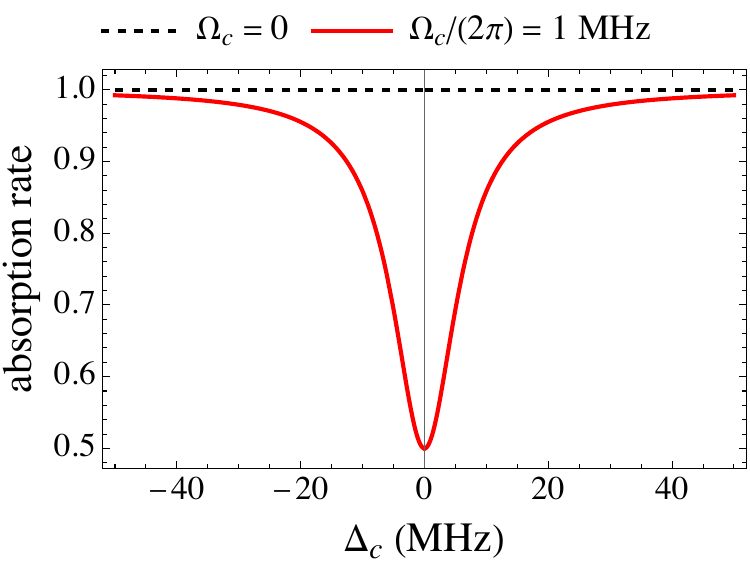}
    \vspace{-4pt}
    \subcaption{The absorption rate of the probe laser.}
  \end{minipage}
  \caption{\small (a) The energy level of EIT system. The probe laser with a frequency $\omega_p$ couples $\ket{1}$ and $\ket{2}$ with a Rabi frequency $\Omega_p$, and the control laser with a frequency $\omega_c$ couples $\ket{2}$ and $\ket{3}$ with a Rabi frequency $\Omega_c$. (b) Plots of the absorption rate of the probe laser. For simplicity, the parameters are set to $\gamma_2/(2\pi)=\gamma_3/(2\pi)=0.1~\si{MHz}$, $\Omega_p/(2\pi)=1~\si{MHz}$ and the resonance condition $\Delta_p=0$ is assumed. The absorption rate is defined as $\Im\chi/(\Im\chi)_{\rm max}$. %The plot shows that the probe laser is completely absorbed when $\Omega_c$ is not present, whereas it becomes transparent when $\Omega_c$ is present.
  }
\vspace{0.6cm}
\end{figure}
%%%%%%%%%%%%%%%%%%
%%%%%%%%%%%%%%%%%%%%%%%%%%%%%%%%%%%%%%%%
\subsection{Electromagnetically induced transparency (EIT)}
In this subsection, we consider the interaction of a three-level system with two electric fields as shown in Fig.\,2 (a) and show the EIT is induced. The probe laser with a frequency $\omega_p$ couples $\ket{1}$ and $\ket{2}$ with the Rabi frequency $\Omega_p$. The control laser with a frequency $\omega_c$ couples $\ket{2}$ and $\ket{3}$ with the Rabi frequency $\Omega_c$. The goal of this subsection is to calculate the density operator for this system and clarify the characteristics of the absorption rate of the probe laser.

The Hamiltonian for this total system is written as
    \begin{eqnarray}
    \hat{H}(t)=\hat{H}_0+\hat{H}'(t)
    \end{eqnarray}
with the unperturbed Hamiltonian
    \begin{eqnarray}
    \hat{H}_0=\varepsilon_1\ket{1}\bra{1}+\varepsilon_2\ket{2}\bra{2}+\varepsilon_3\ket{3}\bra{3}
    \end{eqnarray}
where $\varepsilon_i$ with $i=1,2,3$ denotes the energy of each level. The dipole interaction Hamiltonian reads
    \begin{eqnarray}
    \hat{H}'(t) =-\frac{\Omega_p}{2}e^{-i\omega_p t}\ket{2}\bra{1}-\frac{\Omega_c}{2}e^{-i\omega_c t}\ket{3}\bra{2}+{\rm h.c.}\ ,
    \end{eqnarray}
where $\Omega_p$ and $\Omega_c$ are Rabi frequencies. It is convenient to transform the Hamiltonian to a rotating frame by using
    \begin{eqnarray}
    \hat{\mathcal{H}} = \hat{U}\hat{H}\hat{U}^{\dag}-i\hat{U}\frac{d}{dt}\hat{U}^{\dag}\ ,
    \end{eqnarray}
with the unitary operator
    \begin{eqnarray}
    \hat{U} &\equiv&\exp\left[\,{i\varepsilon_1\ket{1}\bra{1}t}+{i\left(\varepsilon_1+\omega_p\right)\ket{2}\bra{2}t}+{i(\varepsilon_1+\omega_p-\omega_c)\ket{3}\bra{3}t}\right]\ .
    \end{eqnarray}
Then, the Hamiltonian in the rotating frame is written by
    \begin{eqnarray}
\hat{\mathcal{H}}&=&\Delta_p\ket{2}\bra{2}+\left(\Delta_p+\Delta_c\right)\ket{3}\bra{3}+\left[-\frac{\Omega_p }{2}\ket{2}\bra{1}-\frac{\Omega_c }{2}\ket{2}\bra{3}+{\rm h.c.}\right]\ ,
    \end{eqnarray}
where we defined detunings $\Delta_p$ and $\Delta_c$ 
expressed by
    \begin{eqnarray}
    \Delta_p&\equiv&(\varepsilon_2-\varepsilon_1)-\omega_p\ , \\
    \Delta_c&\equiv&(\varepsilon_2-\varepsilon_3)-\omega_c\ .
    \end{eqnarray}
The master equation for the density matrix of the total system in the rotating frame $\tilde{\rho}$ is given by
    \begin{eqnarray}
    \label{master3}
    \frac{d}{dt}\tilde{\rho} = -i[\hat{\mathcal{H}},\tilde{\rho}] +\sum_{k=2,3}\gamma_k\left[\hat{O}_k\tilde{\rho}\hat{O}_k^\dag-\frac{1}{2}\left\{\hat{O}_k^\dag\hat{O}_k,\tilde{\rho}\right\}\right]\ .
    \end{eqnarray}
where $\hat{O}_k=\ket{1}\bra{k}$ represents the relaxation process of the atom. We assume that the atom is in the ground state initially, that is, the initial condition becomes 
$\tilde{\rho}_{11}(0)=1$, $\tilde{\rho}_{12}(0)=\tilde{\rho}_{13}(0)=\tilde{\rho}_{22}(0)=\tilde{\rho}_{23}(0)=\tilde{\rho}_{33}(0)=0$. In order to solve the  master equation~(\ref{master3}), we use linear approximation under the assumption that the frequency of the Rabi oscillation $\Omega_p$ is small enough. 
Under this assumption, we can assume
$\tilde{\rho}_{11}=\Omega_c={\cal O}(1),~ \tilde{\rho}_{12}=\tilde{\rho}_{13}=\tilde{\rho}_{22}=\tilde{\rho}_{23}={\cal O}(\Omega_p)$. 
 Thus, the master equations become
    \begin{eqnarray}
    \frac{d}{dt}\tilde{\rho}_{21}&=&-\left(\frac{\gamma_2}{2}+i\Delta_p\right) \tilde{\rho}_{21}+i\frac{\Omega_p}{2}\tilde{\rho}_{11}+i\frac{\Omega_c}{2}\tilde{\rho}_{31}\ ,\\[6pt]
    \frac{d}{dt}\tilde{\rho}_{31}&=&-\left(\frac{\gamma_3}{2}+i(\Delta_p-\Delta_c)\right)\tilde{\rho}_{31}+i\frac{\Omega_c}{2}\tilde{\rho}_{21}\ .
    \end{eqnarray}
Now, we obtain the stationary state solution $\tilde{\rho}_{21}^{\rm st}$ normalized by $\tilde{\rho}_{11}$ as
    \begin{eqnarray}
	\frac{\tilde{\rho}_{21}^{\rm st}}{\tilde{\rho}_{11}}=\frac{i\left[\gamma_3+2i\left(\Delta_p-\Delta_c\right)\right]\Omega_p}{2\left(\Omega_c^2+\left[\gamma_3+2i\left(\Delta_p-\Delta_c\right)\right]\left[\gamma_2+2i\Delta_p\right]\right)}\ .
	\end{eqnarray}

The result $\tilde{\rho}_{21}^{\rm st}$ in the rotating frame has to be transformed back to $\rho_{21}(t)$ in the original frame by using $\rho_{21}(t)=\tilde{\rho}_{21}(\omega_p)e^{-i\omega_pt}$. Then substituting the $\rho_{21}(t)$ into Eq.~(\ref{chirho}), we finally obtain the absorption rate in terms of the imaginary part of susceptibility and we plotted it as a function of the detuning $\Delta_c$ in Fig.\,2 (b). We see that the absorption rate of the probe laser decreases once the Rabi oscillation between the level $|2\rangle$ and $|3\rangle$ is induced ($\Delta_c\rightarrow 0$) by the control laser. This means that the Rabi oscillation with frequency $\Omega_c\equiv|\bm{d}_{23}\cdot\bm{E}_c|$ between $\ket{2}$ and $\ket{3}$ driven by the control laser $\bm{E}_c\cos\left(\omega_ct\right)$ can interfere with  the Rabi oscillation between $\ket{1}$ and $\ket{2}$ and then the Rydberg atoms become effectively transparent (no absorption) for the probe laser. This situation where the probe laser can pass through the Rydberg atoms without being absorbed is called Electromagnetically Induced Transparency (EIT). 

The point of the EIT is that the electric field, which couples $\ket{2}$ and $\ket{3}$, significantly alters the absorption rate of the probe laser.
Thus, by measuring the changes in the absorption rate of the probe laser, the presence of the electric field that couples $\ket{2}$ and $\ket{3}$ can be detected. This method is further developed into superheterodyne detection method by using a four-level system, which will be explained in the next section.
%%%%%%%%%%%%%%%%%%%%%%
\section{Gravitational wave detector using Rydberg atoms}
%%%%%%%%%%%%%%%%%%%%%%%%%%%%%%%%%%%%%
\begin{figure}[t]
        \vspace{6pt}
        \centering
        \includegraphics[width=11cm]{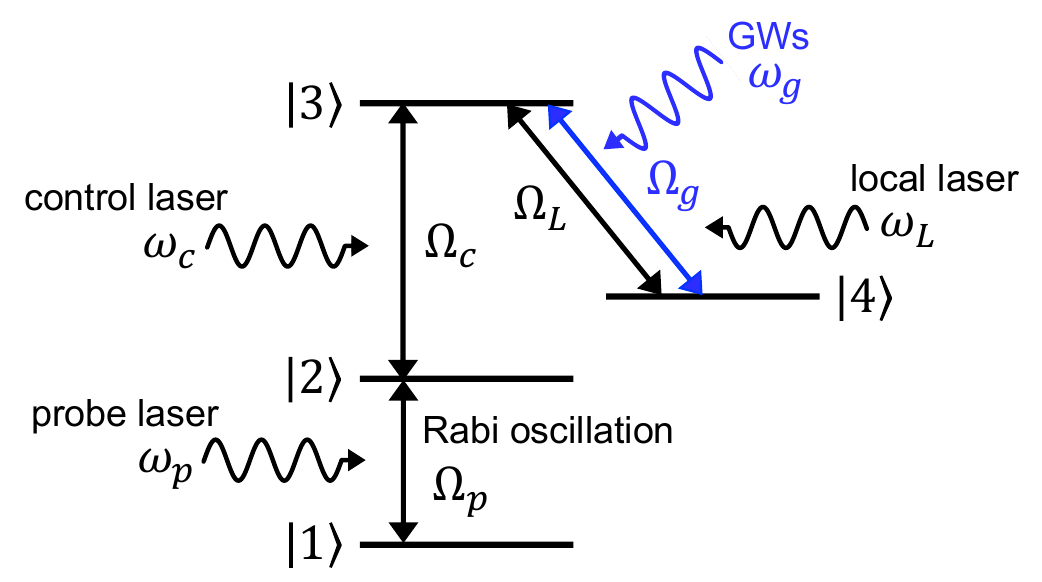}
        \vspace{-2pt}
        \caption{\small Energy levels of the superheterodyne system. The probe laser  induces a Rabi oscillation with a frequency $\Omega_p$  between $\ket{1}$ and $\ket{2}$, the control laser induces Rabi oscillation with frequency $\Omega_p$ between $\ket{2}$ and $\ket{3}$, and the local laser induces Rabi oscillation with frequency $\Omega_L$ between $\ket{3}$ and $\ket{4}$. The electric field induced by GWs with frequency $\omega_g$ in the presence of the magnetic field affects the Rabi oscillation  between $\ket{3}$ and $\ket{4}$.}
        \label{fig:4level}
        \vspace{0.4cm}
\end{figure}
%%%%%%%%%%%%%%%%%%%%%%%%%%%%%%%%%%%%%
\subsection{Superheterodyne detection strategy}
Recently, the authors in~\cite{jing2020atomic} proposed a new method to improve the sensitivity further by combining an electromagnetically induced transparency (EIT) method with superheterodyne detection. This superheterodyne detection utilizes four-level Rydberg atoms, each of which consists of two low energy states $\ket{1}$, $\ket{2}$ and two Rydberg states $\ket{3}$, $\ket{4}$ as depicted in Fig.\,\ref{fig:4level}. In the superheterodyne method, another transition between $\ket{3}$ and $\ket{4}$ is driven by the strong local laser $\bm{E}_L\cos\left(\omega_Lt\right)$.  In this case, the Rabi frequency is given by $\Omega_L\equiv|\bm{d}_{34}\cdot\bm{E}_L|$. The point here is that this local strong laser induces splitting of the EIT peak called Autler-Townes splitting~\cite{Autler:1955,Fleischhauer:2005zz}, that is, splitting of the peak of absorption rate for the probe laser as we will see in Fig.\,\ref{fig:trans}. 

Now, let us suppose that GWs arrive to the Rydberg atoms. If we set up a constant magnetic field away from the four-level Rydberg atoms, then a weak electric field signal with frequency $\omega_g$ is generated from the interaction between the GWs and the magnetic field. This electric field induces the Rabi oscillation with the frequency $\Omega_g\equiv|\bm{d}_{34}\cdot\bm{E}_g|$ between states $\ket{3}$ and $\ket{4}$. As a result, the split absorption rate changes. And this change becomes larger by controlling the local laser as we will see in Eq.~(\ref{kappa0and1}). By splitting the peak of absorption rate of the probe laser in the superhetrodyne method, the sensitivity of the signal of the GW turns out to be improved.
%==========================================================
By using Eq.~(\ref{inducedE}), the Rabi frequency induced by the GWs is given by 
    \begin{eqnarray}
    \label{Omega_g}
    \Omega_g = |\bm{d}_{34}\cdot\bm{E}|=|{d_{34}}^i\varepsilon^{ijk}h^0{}_jB^k|\ ,
    \end{eqnarray}
%where we used the relation (\ref{inducedE}).
Note that $\Omega_g\ll\Omega_L$. This induces a change in the absorption rate of the probe laser when the GWs arrive. More precisely, the frequency shift between the two split peaks can be measured  upon the arrival of GWs. In this way, a signal of the GWs can be measured through the change of the absorption rate of the probe laser. We also note that this superheterodyne method is proposed by the authors in~\cite{Engelhardt:2023qjf} for the detection of axion fields.
%==========================================================
%==========================================================
\subsection{Fermi-normal coordinates}
%In the formula (\ref{Omega_g}),  there appeared the metric perturbations which is gauge dependent. Hence, we need to specify an appropriate coordinate system. 
In Section 2, we considered a constant magnetic field in the laboratory frame. In order to measure GWs that appeared in Eq.~(\ref{Omega_g}) while maintaining the constant magnetic field, it is necessary to introduce a local inertial system for the GWs. We use Fermi-normal (FN) coordinates that describe the effect of gravity from the point of view of an observer in the laboratory. In the FN coordinate, the metric is perturbatively expanded under the condition that the wavelength of the GWs, $\lambda_g$, is much longer than the size of detector $L$, that is, $\lambda_g\gg L$. However, since the detector 
becomes the most sensitive for $\lambda_g\sim L$, we need to improve the FN coordinates so that they can incorporate shorter wavelengths of the GWs. The authors in ~\cite{fortini1982fermi,Marzlin:1994ia,Rakhmanov:2014noa} made this possible in the case that the GW is a plane wave. Recently, the authors in~\cite{Berlin:2021txa} made use of the improved FN coordinate to detect high-frequency GWs with microwave cavities. It is also shown that the sensitivity of magnon GW detectors can be improved by using the FN coordinates~\cite{Ito:2022rxn}. If we assume that the GW is described by the plane wave $h \propto e^{-i(\omega_gt-\bm{k}\cdot\bm{x})}$, the metric components in the FN coordinates are found to be
    \begin{eqnarray}
    g_{00} &=& -1-{R}_{0i0j}(0)x^ix^j \times 2\Re\left[-\frac{i}{\bm{k}\cdot\bm{x}}+\frac{1-e^{-i\bm{k}\cdot\bm{x}}}{(\bm{k}\cdot\bm{x})^2}\right]\,,\\[5pt]
    g_{0i} &=& -\frac{2}{3}{R}_{0jik}(0)x^jx^k \times 3\Re\left[-\frac{i}{2(\bm{k}\cdot\bm{x})}-\frac{e^{-i\bm{k}\cdot\bm{x}}}{(\bm{k}\cdot\bm{x})^2}-i\frac{1-e^{-i\bm{k}\cdot\bm{x}}}{(\bm{k}\cdot\bm{x})^3}\right]\,,\label{g0i}\\[5pt]
    g_{ij} &=& \delta_{ij}-\frac{1}{3}{R}_{ikjl}(0)x^kx^l \times 6\Re\left[-\frac{1+e^{-i\bm{k}\cdot\bm{x}}}{(-\bm{k}\cdot\bm{x})^2}-2i\frac{1-e^{-i\bm{k}\cdot\bm{x}}}{(\bm{k}\cdot\bm{x})^3}\right]\,,
    \end{eqnarray}
where $x^i$ are spatial coordinates, $k^ix_i=\bm{k}\cdot\bm{x}$ and $R_{\mu\nu\rho\sigma}(0)$ is the Riemann tensor evaluated at the origin $x^{i}=0$. In order to evaluate Eq.~(\ref{inducedE}), we focus on Eq.~(\ref{g0i}). 
Since the Riemann tensor in Eq.~(\ref{g0i}) is gauge invariant at the linear order, we can express it by the metric in the transverse-traceless (TT) gauge. Then, Eq.~(\ref{g0i}) in the TT gauge, $h_{0i}$, is calculated as
    \begin{eqnarray}
    h_{0i}&=&\omega_g\left(\left.k_k h_{ji}^{\rm TT}\right|_{\bm{x}=0}x^jx^k-\left.k_i h_{jk}^{\rm TT}\right|_{\bm{x}=0}x^jx^k\right)\left[\frac{\cos(\bm{k}\cdot\bm{x})}{(\bm{k}\cdot\bm{x})^2}-\frac{\sin(\bm{k}\cdot\bm{x})}{(\bm{k}\cdot\bm{x})^3}\right]\,.
    \label{h0i}
    \end{eqnarray}

\begin{figure}[t]
        \centering
        \includegraphics[width=9cm]{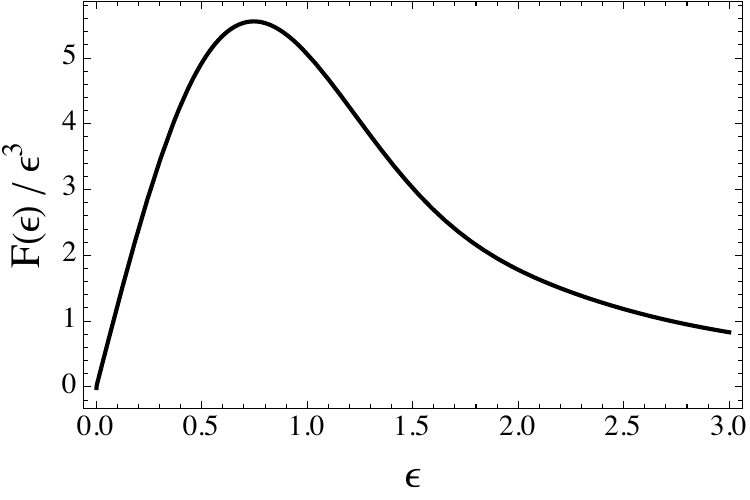}
        \vspace{-4pt}
        \caption{\small Plot of  $F(\epsilon)/\epsilon^3$ as a function of $\epsilon=\ell/\lambda_g=k\ell/(2\pi)=\omega_g\ell/(2\pi)$, which  takes the maximum at $\epsilon\sim 0.7$. Therefore, when the ratio of cell size to gravitational wave wavelength is around 0.7, the Rabi frequency is maximized, leading to the highest sensitivity. Adjusting the size of the cell to match the target gravitational wave wavelength optimizes sensitivity.}
        \label{fig:F(e)}
\end{figure}

Let us assume that the uniform magnetic field $\bm{B}$ is pointing only in the positive $z$ direction. Without loss of generality, we can consider the GWs propagating in the $z$-$x$ plane and its wave vector is given by $\bm{k}=k(\sin\theta, 0, \cos\theta)$. 
Then, the polarization tensors $e^{(+)}_{ij}, e^{(\times)}_{ij}$ are expressed as
    \begin{eqnarray}
	e^{(+)}_{ij}=\frac{1}{\sqrt{2}}
	\begin{pmatrix} 
		  \cos^2\theta & 0 & -\cos\theta \sin\theta \\
		  0 & -1 & 0  \\
		  -\cos\theta \sin\theta & 0 & \sin^2\theta  \\
	\end{pmatrix},
	\hspace{3mm}
	e^{(\times)}_{ij}=\frac{1}{\sqrt{2}}
	\begin{pmatrix} 
		  0 & \cos\theta & 0  \\
		  \cos\theta & 0 & -\sin\theta  \\
		  0 & -\sin\theta & 0 \\
	\end{pmatrix}.
	\end{eqnarray}
The GWs are then expanded in the form of circularly polarized monochromatic plane-wave,
    \begin{eqnarray}
    \label{decomposed}
    h_{ij}^{\rm TT}(t,\bm{x})=h^{(+)}e^{(+)}_{ij}\cos(\omega t -\bm{k}\cdot \bm{x})+h^{(\times)}e^{(\times)}_{ij}\cos(\omega t -\bm{k}\cdot \bm{x})\,,
    \end{eqnarray}
where $h^{(+)}$ and $h^{(\times)}$ are the amplitude of the GWs. Substituting Eq.~(\ref{decomposed}) into Eq.~(\ref{h0i}), we find
    \begin{eqnarray}
    \label{h0x}
    h_{0x}&=&\omega_gk\frac{h^{(+)}}{\sqrt{2}}\sin\theta\left(y^2- z^2 \right)\left[\frac{\cos(\bm{k}\cdot\bm{x})}{(\bm{k}\cdot\bm{x})^2}-\frac{\sin(\bm{k}\cdot\bm{x})}{(\bm{k}\cdot\bm{x})^3}\right]\ ,\\[6pt]
    \label{h0y}
    h_{0y}&=&\omega_gk\frac{h^{(\times)}}{\sqrt{2}}\cos\theta\sin\theta\left(x^2- z^2 \right)\left[\frac{\cos(\bm{k}\cdot\bm{x})}{(\bm{k}\cdot\bm{x})^2}-\frac{\sin(\bm{k}\cdot\bm{x})}{(\bm{k}\cdot\bm{x})^3}\right]\ ,
    \end{eqnarray}
where we ignored the terms involving $x^jx^k=xy,yz,zx$ in Eq.~(\ref{h0i}) because they vanish when 
%vanish by performing triple integrals in spherical coordinates in order to apply 
averaging $h_{0i}$ over the size of detector. Suppose that the size of the detector consisting of the Rydberg atoms is $L$. We average Eqs.~(\ref{h0x}) and ($\ref{h0y}$) over the size of detector radius $\ell=L/2$ in spherical coordinates $(r, \zeta, \phi)$. The details of the calculations are presented in the Appendix~\ref{appA} where we introduced a dimensionless parameter $\epsilon=\ell/\lambda_g=k\ell/(2\pi)$ and a variable $r'=r/\ell$. The spherically-averaged $h_{0i}$ over the size of detector is given in the form
    \begin{eqnarray}
    \label{h0xave}
    \langle h_{0x}\rangle
    &=&\frac{h^{(+)}}{\sqrt{2}}\sin\theta F(\epsilon)\ ,\\[6pt]
    \langle h_{0y}\rangle
    &=&\frac{h^{(\times)}}{\sqrt{2}}\cos\theta\sin\theta F(\epsilon)\ ,
    \label{h0yave}
    \end{eqnarray}
where we used $\omega_g = k$ and defined $F(\epsilon)$ as
\begin{eqnarray}
F(\epsilon)\equiv\frac{\pi\left(\pi^2\epsilon ^2-3\right){\rm Si}(2\pi\epsilon)}{ 2\pi\epsilon}+\frac{\pi\left(2\pi^2\epsilon^2-15\right)\cos (2\pi\epsilon)}{8\pi^2\epsilon^2}+\frac{\pi\left(2\pi^2\epsilon^2+15\right) \sin (2\pi\epsilon )}{16\pi^3\epsilon^3}\ .
\end{eqnarray}
Here, ${\rm Si}(\epsilon)$ is the sine integral defined as ${\rm Si}(x)=\int_0^x dt \sin t/t$. We plotted $F(\epsilon)/\epsilon^3$ as a function of $\epsilon$ in Fig.~\ref{fig:F(e)}. Finally, the averaged Rabi frequency $\Omega_g$ is calculated such as
    \begin{eqnarray}
    \langle \Omega_g\rangle =\left\langle|\bm{d}^{(34)}\cdot\bm{E}|\right\rangle
    &=&|{d^{(34)}}^i\varepsilon^{ijk}\langle h^0{}_j\rangle B^k|\no\\[6pt]
%    &=&-F(\epsilon)\frac{B_z}{\sqrt{2}}\sin\theta\left[d^{(34)}_x h^{(\times)}\cos\theta-d^{(34)}_{y}h^{(+)}\right]\no\\[6pt]
    &=&F(\epsilon)\frac{B_z}{\sqrt{2}}\left|\sin\theta\left[d^{(34)}_x h^{(\times)}\cos\theta-d^{(34)}_{y}h^{(+)}\right]\right|\ .
    \label{omega_g_ave}
    \end{eqnarray}
%%%%%%%%%%%%%%%%%%%%%%%%%%%%%%%%%%%%
%%%%%%%%%%%%%%%%%%%%%%%%%%%%%%%%%%%%

For long-wavelength GWs ($\epsilon\ll 1$), the sensitivity of the detector to the GWs is significantly reduced due to the rapid decay of $F(\epsilon)$ as shown in Fig.~\ref{fig:F(e)}. This reflects the equivalence principle for the detector. On the other hand, for short-wavelength GWs ($\epsilon\gg 1$), $F(\epsilon)$ increases as $F(\epsilon)\propto \epsilon$. However,  as the wavelength of the GWs becomes shorter, the number of Rydberg atoms within the wavelength decreases at a rate proportional to its volume $\epsilon^{-3}$.  Hence, the sensitivity is reduced by $\epsilon^{-3}$.  As a result, the sensitivity of the detector is maximized at $\epsilon\sim 0.7$.

%==========================================================
%==========================================================
%==========================================================
\section{The minimum detectable amplitude of GWs}
In this section, we evaluate the sensitivity of the GW detector with heterodyne receivers. A signal of the GWs is measured by a change in the absorption rate of the probe laser. As mentioned in Sec. 3.2, the absorption rate is proportional to the imaginary part of electric susceptibility which can be related to the density operator of an atom. Thus, we calculate the density operator for the Rydberg four-level system, and then, we evaluate the minimum detectable amplitude of GWs.
%=========================================================
\subsection{Master equation for  Rydberg system}
As we explained in the subsection.~4.1, we consider a four-level Rydberg atom that consists of two low energy states $|1\rangle$, $|2\rangle$ and two Rydberg states $|3\rangle$, $|4\rangle$ as depicted in Fig.\,\ref{fig:4level}.
The total Hamiltonian $\hat{H}(t)$ for the four-level atom interacting with probe laser, control laser, strong local laser and GWs is written as
\begin{eqnarray}
\label{H(t)}
\hat{H}(t)&=&\hat{H}_0 + \hat{H}'(t)\ ,
\end{eqnarray}
with the unperturbed Hamiltonian  
\begin{eqnarray}
\hspace{-1cm}\hat{H}_0 =\varepsilon_1\ket{1}\bra{1}+\varepsilon_2\ket{2}\bra{2}+\varepsilon_3\ket{3}\bra{3}+\varepsilon_4\ket{4}\bra{4}\ ,
\end{eqnarray}
where $\varepsilon_i$ with $i=1,2,3,4$ denotes the energy of each level. The dipole interaction Hamiltonian reads
\begin{eqnarray}
\hat{H}'(t) =-\frac{\Omega_p}{2}e^{-i\omega_p t}\ket{2}\bra{1}-\frac{\Omega_c}{2}e^{-i\omega_c t}\ket{3}\bra{2}-\left(\frac{\Omega_L}{2}e^{-i\omega_L t}+\frac{\Omega_g}{2}e^{-i\omega_g t}\right)\ket{3}\bra{4}+{\rm h.c.}\ ,
\end{eqnarray}
where $\Omega_p$, $\Omega_c$, $\Omega_L$, and $\Omega_g$, are Rabi frequencies associated with the transitions $\ket{1}\rightarrow\ket{2}$, $\ket{2}\rightarrow\ket{3}$ and $\ket{3}\rightarrow\ket{4}$, respectively. Using
 the unitary operator
\begin{eqnarray}
\hat{U} &\equiv&\exp\Bigl[\,{i\varepsilon_1\ket{1}\bra{1}t}+{i\left(\varepsilon_1+\omega_p\right)\ket{2}\bra{2}t}\Bigr.\no\\[6pt]
    &&\hspace{1.5cm}\Bigl.+{i(\varepsilon_1+\omega_p+\omega_c)\ket{3}\bra{3}t}+{i(\varepsilon_1+\omega_p+\omega_c-\omega_{L})\ket{4}\bra{4}t}\,\Bigr]\ ,
    \end{eqnarray}
we obtain the Hamiltonian in the rotating frame 
    \begin{eqnarray}
    \hat{\mathcal{H}}&=&\Delta_p\ket{2}\bra{2}+\left(\Delta_p+\Delta_c\right)\ket{3}\bra{3}+\left(\Delta_p+\Delta_c+\Delta_L\right)\ket{4}\bra{4}\no\\[6pt]
    &&\hspace{2cm}+\left[-\frac{\Omega_p }{2}\ket{2}\bra{1}-\frac{\Omega_c }{2}\ket{3}\bra{2}-\frac{\Omega_L+\Omega_g e^{-i\delta_g t}}{2}\ket{3}\bra{4}+{\rm h.c.}\right]\ ,
    \end{eqnarray}
where we defined detunings  $\Delta_p$, $\Delta_c$, $\Delta_L$, and $\delta_g$ by
    \begin{eqnarray}
    \Delta_p&\equiv&(\varepsilon_2-\varepsilon_1)-\omega_p\ , \\
    \Delta_c&\equiv&(\varepsilon_3-\varepsilon_2)-\omega_c\ , \\
    \Delta_L&\equiv&\omega_L-(\varepsilon_3-\varepsilon_4)\ , \\
    \delta_g&\equiv&\omega_g-\omega_L\ .
    \end{eqnarray}
Note that the time dependence except for $\delta_g$ in the Hamiltonian was eliminated in this rotating frame. The master equation for total system in the rotating frame $\tilde{\rho}$ is given by
    \begin{eqnarray}
    \label{master}
    \frac{d}{dt}\tilde{\rho} &=& -i[\hat{\mathcal{H}},\tilde{\rho}] +\sum_{k=2,3,4}\gamma_k\left[\hat{O}_k\tilde{\rho}\hat{O}_k^\dag-\frac{1}{2}\left\{\hat{O}_k^\dag\hat{O}_k,\tilde{\rho}\right\}\right]\no\\[6pt]
    &&\hspace{40pt}+\left(\gamma_2\tilde{\rho}_{22}+\gamma_{4}\tilde{\rho}_{44}\right)\ket{1}\bra{1}+\gamma_3\tilde{\rho}_{33}\ket{2}\bra{2}\ ,
    \end{eqnarray}
where $\hat{O}_k=\ket{1}\bra{k}$ represents the relaxation process of the atom, and the third and the fourth term in the right hand side represent repopulation processes.

Let us consider the resonance case $\Delta_L=0$. We can assume that $\gamma_3=\gamma_4=0$ because the relations $\gamma_3, \gamma_4 \ll \gamma_2$ hold in general. Then, the stationary state solution of the  master equation $\tilde{\rho}_{21}^{\rm st}$ is
    \begin{eqnarray}
     \frac{\tilde{\rho}_{21}}{\tilde{\rho}_{11}}&=&\frac{\left[4\left(\Delta_p+\Delta_c\right)^2-\Omega^2\right]\left[-8\Delta_p\left(\Delta_p+\Delta_c\right)^2+2\left(\Delta_p+\Delta_c\right)\Omega_c^2+2\Delta_p\Omega^2\right]\Omega_p}{A'\Delta_c^4+B'\Delta_c^3+C'\Delta_c^2+D'\Delta_c+E'}\no\\[6pt]
     &&\h +i \frac{\gamma_2 \left[4\left(\Delta_p+\Delta_c\right)^2-\Omega^2\right]^2\Omega_p}{A'\Delta_c^4+B'\Delta_c^3+C'\Delta_c^2+D'\Delta_c+E'}\ ,
    \end{eqnarray}
where we defined $A$, $B$, $C$, $D$, and $E$ as
    \begin{eqnarray}
    A'\hspace{-5pt}&=&\hspace{-5pt}16\left(4\Delta_p^2+\Omega_p^2+\gamma_2^2\right)\ ,\no\\[4pt]
    B'\hspace{-5pt}&=&\hspace{-5pt}256\Delta_p^3+32\Delta_p\left(2\Omega_p^2-\Omega_c^2+2\gamma_2^2\right)\ ,\no\\[4pt]
    C'\hspace{-5pt}&=&\hspace{-5pt}384\Delta_p^4+32\Delta_p^2\left(3\Omega_p^2-3\Omega_c^2-\Omega^2+3\gamma_2^2\right)+4\left(\Omega_p^2\Omega_c^2-2\Omega_p^2\Omega^2+\Omega_c^4-2\Omega^2\gamma_2^2\right)\ ,\no\\[4pt]
    D'\hspace{-5pt}&=&\hspace{-5pt}256\Delta_p^5+32\Delta_p^3\left(2\Omega_p^2-3\Omega_c^2-2\Omega^2+2\gamma_2^2\right)+8\Delta_p\left(\Omega_p^2\Omega_c^2-2\Omega_p^2\Omega^2+\Omega_c^4+\Omega_c^2\Omega^2-2\Omega^2\gamma_2^2\right)\ ,\no\\[4pt]
    E'\hspace{-5pt}&=&\hspace{-5pt}64\Delta_p^6+16\Delta_p^4\left(\Omega_p^2-2\Omega_c^2-2\Omega^2+\gamma_2^2\right)\no\\
    &&+4\Delta_p^2\left(\Omega_p^2\Omega_c^2-2\Omega_p^2\Omega^2+\Omega_c^4+2\Omega_c^2\Omega^2+\Omega^4-2\Omega^2\gamma_2^2\right)+\Omega_p^2\Omega_c^2\Omega^2+\Omega_p^2\Omega^4+\Omega^4\gamma_2^2\ .
    \end{eqnarray}
Here we introduced $\Omega\equiv\Omega_L+\Omega_g e^{-i\delta_g t}$ which is almost constant because of the relation $\Omega_g\ll\Omega_L$.

%%%%%%%%%%%%%%%%%%%%%%%%%%%%%%%%%%%%
\begin{figure}[t]
        \centering
        \includegraphics[width=8.5cm]{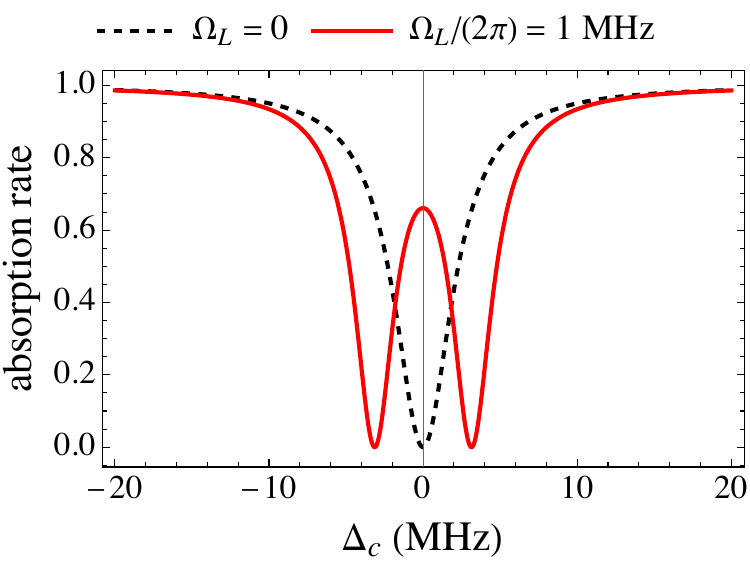}
        \vspace{-2pt}
        \caption{\small Plots of the absorption rate of the probe laser as a function of $\Delta_c$. Here we used $\gamma_2/(2\pi)=5.2~\si{MHz}$, $\gamma_3/(2\pi)=3.9~\si{kHz}$, $\gamma_4/(2\pi)=1.7~\si{kHz}$, $\Omega_p/(2\pi)=5.7~\si{MHz}$, $\Omega_c/(2\pi)=0.97~\si{MHz}$, $\Omega_L/(2\pi)=1.0~\si{MHz}$~\cite{jing2020atomic}. The absorption rate is defined as $\Im\chi/(\Im\chi)_{\rm max}$. The splitting of the peak of absorption rate for the probe laser occurs due to the local laser.}
        \label{fig:trans}
        \vspace{0.6cm}
\end{figure}
%%%%%%%%%%%%%%%%%%%%%%%%%%%%%%%%%%%%

The result $\tilde{\rho}_{21}^{\rm st}$ in the rotating frame has to be transformed back to $\rho_{21}(t)$ in the original frame by using $\rho_{21}(t)=\tilde{\rho}_{21}(\omega_p)e^{-i\omega_pt}$. Substituting $\rho_{21}(t)$ into Eq.~(\ref{chirho}), we finally obtain the absorption rate in terms of the imaginary part of susceptibility in the form.
    \begin{eqnarray}
    \label{solchi}
    \Im\chi=\frac{2N|\bm{d}_{12}|^2}{\Omega_p}\frac{\gamma_2 \left[4\left(\Delta_p+\Delta_c\right)^2-\Omega^2\right]}{A\Delta_c^4+B\Delta_c^3+C\Delta_c^2+D\Delta_c+E}\ .
    \end{eqnarray}
The absorption rate, defined as $\Im \chi/(\Im \chi)_{\rm max}$, is shown in Figure \ref{fig:trans}. It can be seen that the local laser causes the AT splitting. This is a phenomenon that does not occur in the three-level EIT system, and is a feature of the supheterodyne system.
%======================================================
\subsection{Estimation of the minimum detectable GW amplitude}
\begin{figure}[t]
        \centering
        \vspace{-2pt}
        \includegraphics[width=8.5cm]{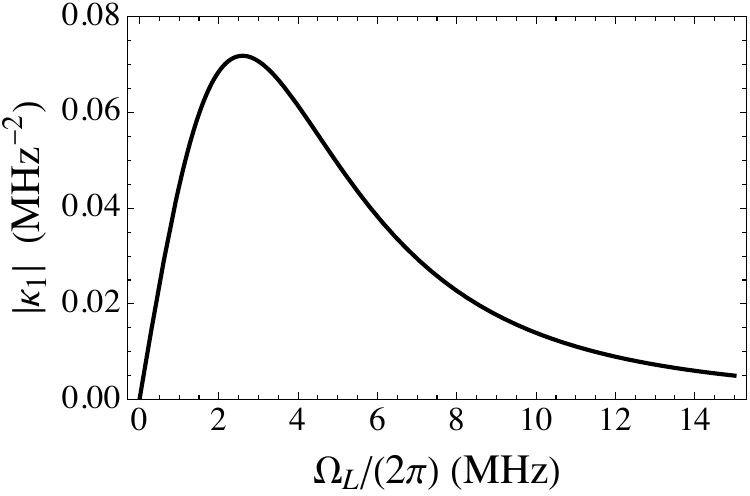}
        \caption{\small Plot of $|\kappa_1|$ as a function of the Rabi frequency $\Omega_L$ normalized by $2\pi$ in which the resonance condition $\Delta_p=\Delta_c=\Delta_L=0$ is imposed. Other parameters are set to $\gamma_2/(2\pi)=5.2~\si{MHz}$, $\gamma_3=\gamma_4=0$, $\Omega_p/(2\pi)=5.7~\si{MHz}$, $\Omega_c/(2\pi)=0.97~\si{MHz}$, $\Omega_L/(2\pi)=1.0~\si{MHz}$. This plot shows that  $|\kappa_1|$ becomes the maximum value  $0.072~\si{MHz^{-2}}$ at the Rabi frequency  $\Omega_L/(2\pi)=2.6~\si{MHz}$. The minimum detectable amplitude of the GWs, associated with these values, can be measured.}
        \label{fig:kappa}
        \vspace{0.6cm}
\end{figure}

To evaluate the minimum detectable amplitude of GWs, we restore the time-dependence of $\Omega$ in Eq.~(\ref{solchi}). We can expand $|\Omega|^2$ up to the first order in $\Omega_g$ such as 
$$
|\Omega^2|=\left(\Omega_L+\Omega_g\,e^{-i\delta_g t}\right)\left(\Omega_L+\Omega_g\,e^{i\delta_g t}\right)
\sim\Omega_L^2+2\Omega_L\Omega_g\cos\delta_gt . 
$$ 
Plugging the result into Eq.~(\ref{solchi}), we find
    \begin{eqnarray}
    \label{kappa0and1}
    \Im\chi&=&2N|\bm{d}_{12}|^2\kappa_0 + 2N|\bm{d}_{12}|^2\kappa_1 \Omega_g \cos(\delta_gt)
    \end{eqnarray}
where $\kappa_0$ and $\kappa_1$ depend on the Rabi frequencies $\Omega_p, \Omega_c, \Omega_L$, the detunings $\Delta_p, \Delta_c$ and relaxation rate $\gamma_2$
and can be calculated as
\begin{eqnarray}
\kappa_0=\frac{1}{\Omega_p}\Im\rho_{12}\left(\Omega_L^2\right), \hhh \kappa_1=2\frac{\Omega_L}{\Omega_p}\frac{d\Im\rho_{21}(x)}{dx}\bigg|_{x=\Omega_L^2}\ .
\label{bothkappa}
\end{eqnarray}
Note that $\kappa_1$ represents a change of absorption rate of the probe laser due to the GWs. %Hence, we need to maximize $\kappa_1$. 
Notice that $\kappa_1$ would not exist without the local laser. We plotted the $\Omega_L$ dependence of $\kappa_1$ in Fig.~\ref{fig:kappa}. We see  $|\kappa_1|$ takes the maximum value $0.072~\si{MHz^{-2}}$ for $\Omega_L/(2\pi)=2.6~\si{MHz}$. 
%%%%%%%%%%%%%%%%%%%%%%%%%%%%%%%%%%%%
%%%%%%%%%%%%%%%%%%%%%%%%%%%%%%%%%%%%

By using Eqs.~(\ref{E}) and (\ref{kappa}), the output power of the probe laser $P(t)$ is calculated as $P(t)=|E\left(z,t\right)|^2=|E_0|^2e^{-2\omega_p\kappa z}=P_ie^{-\omega_pL \Im\chi}$, where $P_i(=|E_0|^2)$ is input power of the probe laser, $L$ is the size of Rydberg atoms and we set $n=1$, $z=L$. Substituting Eq.~(\ref{kappa0and1}) into $P(t)$, we find
    \begin{eqnarray}
    P(t)&=&P_ie^{-\omega_pL 2N|\bm{d}_{12}|^2\kappa_0}e^{-\omega_pL 2N|\bm{d}_{12}|^2\kappa_1\Omega_g \cos(\omega_gt)}\no\\[6pt]
    &\approx&P_ie^{-\omega_pL 2N|\bm{d}_{12}|^2\kappa_0}-P_ie^{-\omega_pL 2N|\bm{d}_{12}|^2\kappa_0}\cdot 2N|\bm{d}_{12}|^2\omega_pL \kappa_1\Omega_g \cos(\omega_gt)\no\\[6pt]
    &\equiv&P_0+2P_0N|\bm{d}_{12}|^2\omega_pL \kappa_1\Omega_g \cos(\delta_gt)\,,
    \label{output}
    \end{eqnarray}
where $P_0\equiv P_ie^{-\omega_pL 2N|\bm{d}_{12}|^2\kappa_0}$. Now, we estimate the minimum detectable electric field. Note that a signal can be measured when the ratio of the stationary term to the oscillation term of the Eq.(3.31) is greater than 1. Thus, the minimum detectable Rabi-frequency $\Omega_g{}_{\rm (min)}$ satisfies the following equation.
\begin{equation}
\frac{2P_0N|\bm{d}_{12}|^2 \omega_pL|\kappa_1| \Omega_g \cos(\delta_g)}{P_0}=1\ .
\end{equation}
Combining it with the relation $\Omega_g{}_{\rm (min)}=|\bm{d}_{34}\cdot \bm{E}_{\rm min}|$, we find  the minimum magnitude of the detectable electric field  as
\begin{equation}
E_{\rm min}=\frac{\Omega_g{}_{\rm (min)}}{|\bm{d}_{34}|}=\frac{1}{2|\bm{d}_{34}||\bm{d}_{12}|^2 N\omega_pL|\kappa_1|}\ .
\end{equation}

Let us estimate the minimum detectable value. We consider Rubidium atoms with the four levels $\ket{1}:5\ce{S}_{1/2}$, $\ket{2}:5\ce{P}_{3/2}$, $\ket{3}:59\ce{P}_{3/2}$, and $\ket{4}:57\ce{D}_{5/2}$. In this case, the dipole moment and the energy gap are calculated as $|\bm{d}_{12}|=5.158ea_0$, $\varepsilon_2-\varepsilon_1=384~\si{THz}$, $|\bm{d}_{34}|=2416ea_0$, and $\varepsilon_3-\varepsilon_4=26.4~\si{GHz}$. Here, $e$ is a electric charge and $a_0$ is the Bohr radius, and we used Alkari Rydberg Calculator (ARC) package~\cite{robertson2021arc} to calculate the dipole moment. According to~\cite{kubler2010coherent}, $N$ can be increased up to $N\sim 1.6\times10^{14}~\si{cm^{-3}}$. The size of the Rydberg atoms is $L=2\ell=10~\si{cm}$. Thus, we obtain
    \begin{eqnarray}
    \frac{E_{\rm min}}{\sqrt{\si{Hz}}}&=&8.8\times10^{-17}~\si{eV^2/\sqrt{\si{Hz}}}\left(\frac{2416~ea_0}{|\bm{d}_{34}|}\right)\left(\frac{5.158~ea_0}{|\bm{d}_{12}|}\right)^2\left(\frac{1.6\times 10^{14}~\si{cm^{-3}}}{N}\right)\no\\[6pt]
    &&\hspace{4.5cm}\times\left(\frac{384~\si{THz}}{\omega_p}\right)\left(\frac{10~\si{cm}}{L}\right)\left(\frac{0.072~\si{MHz^{-2}}}{|\kappa_1|}\right)\,.
\end{eqnarray}
Thus, the minimum detectable electric fields turn out to be $E_{\rm min}=8.8\times10^{-17}~\si{eV^2/\sqrt{\si{Hz}}}=1.4~\si{pV/(cm\cdot\sqrt{\si{Hz}})}$. 

Next we estimate the minimum detectable amplitude of GWs. In Eq.~(\ref{omega_g_ave}), to consider the most favourable situation, we choose $\theta=\pi/2$ that represents the situation where GWs propagate in the direction perpendicular to the magnetic field. In this case, 
from Eq.~(\ref{omega_g_ave}), $h$ is expressed as
    \begin{eqnarray}
    \frac{h_{\rm min}}{\sqrt{\si{Hz}}}
    = \frac{1}{\sqrt{\si{Hz}}}\frac{\sqrt{2}\,\langle \Omega_g\rangle}{|\bm{d}_{34}|B_zF(\epsilon)}
    = \frac{1}{\sqrt{\si{Hz}}}\frac{\sqrt{2}}{2}\frac{1}{B_zF(\epsilon)|\bm{d}_{34}||\bm{d}_{12}|^2N\omega_p\,L\,|\kappa_1|}\ .
    \end{eqnarray}
Here, we take $B_z=10~\si{T}$. The energy gap of the two Rydberg states determines the detectable frequency of the GWs,  $\omega_g\sim\varepsilon_4-\varepsilon_3=26.4~\si{GHz}$, which corresponds to the wavelength $\lambda_g=1.14$ cm. When the size of the Rydberg atoms is $L=2\ell=10~\si{cm}$, then the value of $F(\epsilon)$ at $\epsilon=\omega_g\ell/2\pi=0.701$ becomes $F(0.701)\simeq 1.91$. Thus, the minimum detectable amplitude of the GWs by using the Rydberg atoms turns out to be
    \begin{eqnarray}
\frac{h_{\rm min}}{\sqrt{\si{Hz}}}
    &=&2.8\times10^{-20} \frac{1}{\sqrt{\si{Hz}}}
    \left(\frac{1.91} 
    {F(\epsilon)}\right)\left(\frac{10~\si{T}}{B_z}\right)\left(\frac{2416~ea_0}{|\bm{d}_{34}|}\right)\left(\frac{5.158~ea_0}{|\bm{d}_{12}|}\right)^2
    \no\\
    &&\hspace{0.8cm}\times 
    \left(\frac{1.6\times 10^{14}~\si{cm^{-3}}}{N}\right)
    \left(\frac{384~\si{THz}}{\omega_p}\right)\left(\frac{10~\si{cm}}{L}\right)\left(\frac{0.072~\si{MHz^{-2}}}{|\kappa_1|}\right)
    \,.
    \label{hmin}
    \end{eqnarray}

Finally, let us estimate the minimum detectable amplitude of GWs in the quantum projection noise limit based on~\cite{fan2015atom}.
The basic idea of the EIT is to measure the frequency shift between the two split peaks $\Delta\nu$
which is related to the amplitude of electric field $E$ and electric dipole moment $d$
as
\begin{eqnarray}
   E \simeq  \frac{\Delta \nu}{d} \ .
\end{eqnarray}
Since $\Delta \nu \sim 1/(\sqrt{n}\,T_2)$, where $T_2$ is the coherence time of the EIT and 
$n$ is the number of independent measurement taking place per second,
we obtain 
\begin{eqnarray}
   E \simeq  \frac{1}{d \sqrt{n} T_2} \ .
\end{eqnarray}
Let $T$ be the integration time of the coherent EIT process.
Then, we can deduce 
\begin{eqnarray}
   n = N_a\frac{T}{T_2} \ ,
\end{eqnarray}
where $N_a$ is the number of atoms participating in the measurement per second. Hence, the minimum detectable electric field in the quantum projection noise limit~(QPNL) is given by
\begin{eqnarray}
   E_{\rm min} =  \frac{1}{d \sqrt{N_a TT_2}} \ .
\end{eqnarray}
Setting $T=1$s gives the sensitivity limit
\begin{eqnarray}
   E_{\rm QPNL} =  \frac{E_{\rm min}}{\sqrt{\rm Hz}} = \frac{1}{d \sqrt{N_a T_2}} \ .
\end{eqnarray}
The dipole moment is $d=|\bm{d}_{34}|=2416 ea_0\sim 3.87\times10^{-6}~\si{cm}$. The number of atoms participating in the measurement is $N_a=2.14\times 10^{13}~\si{s^{-1}}$~\cite{jing2020atomic}, and the coherence time of Rydberg atom EIT system is $T_2=100~\si{ns}$~\cite{kubler2010coherent}. Thus, we obtain
\begin{eqnarray}
   E_{\rm QPNL} &=& 1.7\times 10^{-1}~\si{pV/(cm\cdot \sqrt{Hz})}\no\\[6pt]
   &&\times\left(\frac{3.87\times10^{-6}\,{\rm cm}}{d}\right)\left(\frac{2.14\times 10^{13}\,{\rm s^{-1}}}{N_a}\right)^{1/2}\left(\frac{100\,{\rm ns}}{T_2}\right)^{1/2}\ .  \quad
\end{eqnarray}
This can be translated to the amplitude of GWs as
\begin{eqnarray}
      h_{\rm QPNL}
    &=&3.4\times10^{-21} \frac{1}{\sqrt{\si{Hz}}}
    \left(\frac{1.91} 
    {F(\epsilon)}\right)\left(\frac{10~\si{T}}{B_z}\right)
    \no\\
    &&\hspace{0.8cm}\times 
   \left(\frac{3.87\times10^{-6}\,{\rm cm}}{d}\right)\left(\frac{2.14\times 10^{13}\,{\rm s^{-1}}}{N_a}\right)^{1/2}\left(\frac{100\,{\rm ns}}{T_2}\right)^{1/2}
    \,.
    \label{hmin}
    \end{eqnarray}
For instance, we can expect ultimate value $1.7~\si{fV/cm}$ for $T=10^4$ s.
This number can be translated into the GW amplitude  $h \sim 10^{-23}$.

We note that one of the noise sources is thermal black-body radiation. However, background electromagnetic waves can be shielded by enclosing  the detector in metal, so they do not become a source of noise that would affect the sensitivity. Although other significant sources of noise include laser amplitude fluctuations and measurement noise, according to~\cite{meyer2020assessment}, shot noises and measurement noises can be neglected. Therefore, the fundamental noise limit is determined by the quantum projection noise limit Eq.~(\ref{hmin}). Note that we are evaluating the minimum detectable GW amplitude in an ideal scenario where mutual interactions between Rydberg atoms are neglected. A future challenge will be to estimate, both experimentally and theoretically, how interactions between atoms affect sensitivity.

%%%%%%%%%%%%%%%%%%%%%%%%%%%%%%%%%%%%
\begin{figure}[H]
        \centering
        \includegraphics[width=11cm]{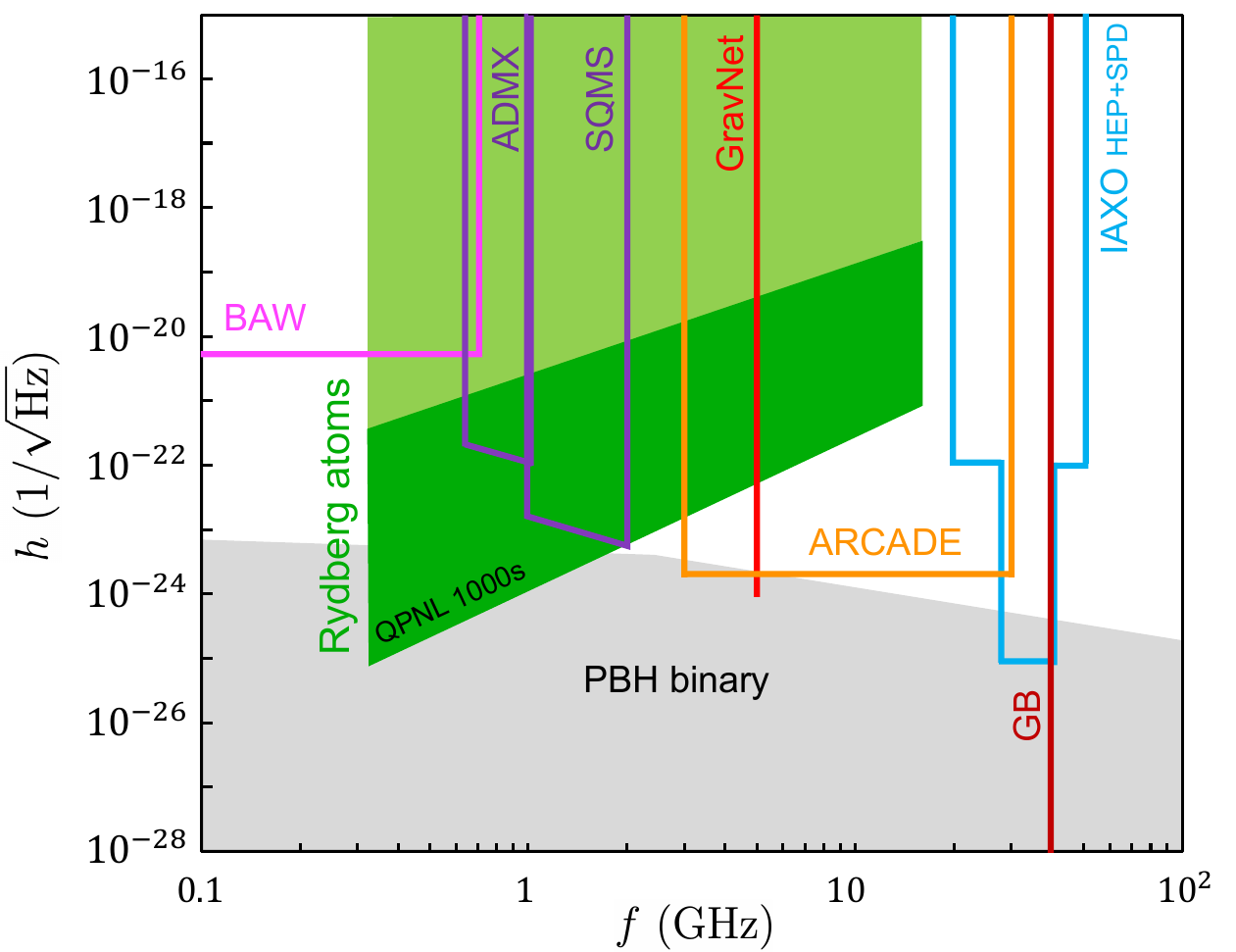}
        \vspace{-2pt}
        \caption{\small An overview of the characteristic strain for gravitational wave detector in the frequency range between $0.1$ and $100$ GHz. Light green region is for the superheterodyne method with Rydberg atoms. Dark green region is for the quantum projection noise limit(QPNL) in a 1000 second measurement with Rydberg atoms. The sensitivity of bulk acoustic wave devices (BAW)~\cite{goryachev2013observation, galliou2013extremely} are shown in light red line. The sensitivity of microwave cavities from ADMX~\cite{ADMX:2021nhd} and SQMS~\cite{Berlin:2021txa} is shown in purple. Global network of cavities(GravNet)~\cite{Schmieden:2023fzn} is shown in red. The sensitivity of the ARCADE 2~\cite{Fixsen:2009xn} is shown in orange, IAXO HEP and IAXO SPD~\cite{Ringwald:2020ist} are shown in light blue, and EM Gaussian beams(GB)~\cite{li2003electromagnetic} is shown in brown. The grey region shows the gravitational waves emitted from the primordial blake hole (PBH) binary~\cite{Franciolini:2022htd}.}
        \label{fig:sensitivity}
        \vspace{6pt}
\end{figure}
%%%%%%%%%%%%%%%%%%%%%%%%%%%%%%%%%%%%

\subsection{Comparison with other detectors}
The detector using Rydberg atoms has an advantage over other GW detectors in the broad range of high-frequencies. For instance, it is possible to adjust the detectable frequencies of GWs by tuning the energy level of $\ket{3}$ and $\ket{4}$ and the size of the cell of Rydberg atoms. We would be able to  prepare multiple cells of Rydberg detectors and set  each one to target gravitational waves at different frequencies. Thus, GW detector by Rydberg atoms make us possible to detect a wide range of frequencies. It is also possible to search for GWs in the angular frequency range between $2$ GHz and $100$ GHz, which corresponds to the frequency range between $0.3$ GHz and $16$ GHz. The lower limit is determined by the minimum energy level gap between $\ket{3}$ and $\ket{4}$, which arises from the fact that the principal quantum number of maximum excitation in a Rydberg atom is $\mathcal{O}(90)$. The upper limit, on the other hand, is determined by the sensitivity. 
At frequencies higher than around $16~\si{GHz}$, the minimum amplitude of the detectable GWs becomes above $10^{-19}$. Hence it would be difficult to detect GWs. Their minimum detectable amplitude can be estimated from Eq.~(\ref{hmin}). The sensitivity and the range of observable frequencies of various detectors are shown in Fig.~\ref{fig:sensitivity} where we can see the Rydberg detector can cover a wide range of GWs.

%==========================================================
\section{Conclusion}
 We studied high-frequency GW detectors with Rydberg atoms. First, we showed that a weak electric field signal is generated from GWs in the presence of the magnetic field. We calculated the effective electric field induced by GWs using improved Fermi-normal coordinates with which we can treat GWs with wavelength shorter than the detector size. We then explained the electromagnetically induced transparency (EIT) employed for detecting electric fields.
 The method is further improved by combining the superheterodyne detection method in the system of Rydberg atoms. The weak signal of the GWs can be probed by measuring the absorption rate of the probe laser. 
We showed that the absorption rate is proportional to the imaginary part of electric susceptibility, and that the susceptibility can be related to the density operator of the Rydberg atom. Considering two low energy states $|1\rangle$, $|2\rangle$ and two Rydberg states $|3\rangle$, $|4\rangle$, we obtain the absorption rate of the probe laser in terms of the imaginary part of the susceptibility. We evaluated the ratio of the output power with a signal of GWs to that with no signal GWs. Finally, we evaluated the the minimum measurable GW amplitude with Rubidium Rydberg atoms and found that GWs with the amplitude $10^{-20}$ and the angular frequency $26.4$ GHz can be detected.

The advantage of using the GW detector with Rydberg atoms is its broad detectable frequency band due to the flexibility of adjusting the detector size. 
Tuning the energy levels of $\ket{3}$ and $\ket{4}$ and the size of the cell of Rydberg atoms, we can detect GWs with a frequency in the range between $0.3$ GHz and $16$ GHz.
 
%==========================================================
\section*{Acknowledgments}
S.\ K. was supported by the Japan Society for the Promotion of Science (JSPS) KAKENHI Grant Numbers JP22K03621, JP22H01220, 24K21548 and MEXT KAKENHI Grant-in-Aid for Transformative
Research Areas A “Extreme Universe” No. 24H00967. J.\ S. was in part supported by JSPS KAKENHI Grant Numbers  JP20H01902 and JP22H01220, JP23K22491, and JP24K21548. A.\ T. was supported by Research Support Scholarship from Kuroda Scholarship Foundation.

%==========================================================
\appendix

%==========================================================
\section{Rabi frequency stemming from GWs}
\label{appA}
When we derive eq.~(\ref{h0xave}) and eq.~(\ref{h0yave}), we perform integration in spherical coordinates $(r, \zeta, \phi)$ as follows.

%==========================================================
For $i=x$, we can calculate as
    \begin{eqnarray}
    k_k h_{ji}^{\rm TT} x^jx^k
    &=&k\sin\theta\left(h_{xx}xx + h_{yx}yx+h_{zx}zx\right) + k\cos\theta\left(h_{xx}xz + h_{yx}yz+h_{zx}zz\right)\no\\[6pt]
    &=&k\sin\theta\left(\frac{h^{(+)}}{\sqrt{2}}\cos^2\theta xx + \frac{h^{(\times)}}{\sqrt{2}}\cos\theta yx - \frac{h^{(+)}}{\sqrt{2}}\cos\theta \sin\theta zx\right)\no\\[6pt]
    &&+ k\cos\theta\left(\frac{h^{(+)}}{\sqrt{2}}\cos^2\theta xz + \frac{h^{(\times)}}{\sqrt{2}}\cos\theta yz - \frac{h^{(+)}}{\sqrt{2}}\cos\theta \sin\theta zz\right)\,,
    \end{eqnarray}
and
    \begin{eqnarray}
    k_i h_{jk}^{\rm TT} x^jx^k&=&
    k_x h_{xx} xx + k_x h_{yy} yy + k_x h_{zz} zz + 2k_x h_{xy}xy + 2k_x h_{yz}yz + 2k_xh_{zx}zx\no\\[6pt]
    &=&k\sin\theta\left(\frac{h^{(+)}}{\sqrt{2}}\cos^2\theta xx - \frac{h^{(+)}}{\sqrt{2}}yy + \frac{h^{(+)}}{\sqrt{2}}\sin^2\theta zz\right.\no\\[6pt]
    &&\hspace{12mm}\left. +2\frac{h^{(\times)}}{\sqrt{2}}\cos\theta xy - 2\frac{h^{(\times)}}{\sqrt{2}}\sin\theta yz - 2\frac{h^{(\times)}}{\sqrt{2}}\cos\theta\sin\theta zx\right) \ .
    \end{eqnarray}
When integrating with respect to $\phi$ in spherical coordinates, terms involving $xy,xz,yz$ will be zero. Thus we obtain
    \begin{eqnarray}
    h_{0x}&=&\omega_gk\left(- \frac{h^{(+)}}{\sqrt{2}}\cos^2\theta \sin\theta z^2 + \frac{h^{(+)}}{\sqrt{2}}\sin\theta y^2 - \frac{h^{(+)}}{\sqrt{2}}\sin^3\theta z^2 \right)\left[\frac{\cos(\bm{k}\cdot\bm{x})}{(\bm{k}\cdot\bm{x})^2}-\frac{\sin(\bm{k}\cdot\bm{x})}{(\bm{k}\cdot\bm{x})^3}\right]\no\,,\\[6pt]
    &=&\omega_gk\frac{h^{(+)}}{\sqrt{2}}\sin\theta\left(y^2- z^2 \right)\left[\frac{\cos(\bm{k}\cdot\bm{x})}{(\bm{k}\cdot\bm{x})^2}-\frac{\sin(\bm{k}\cdot\bm{x})}{(\bm{k}\cdot\bm{x})^3}\right]\,.
    \end{eqnarray}
We align the $z$-axis with the direction of $\bm{k}$ and introduce a spherical coordinate system. We then convert the Cartesian coordinate to the spherical coordinate by using   $x=r\sin\zeta\cos\phi\,,~y=r\sin\zeta\sin\phi\,,~z=r\cos\zeta$. 
Using $\omega_g = k$, 
$\bar{\epsilon}=k\ell, r'=r/\ell$, 
we can take average
    \begin{eqnarray}
    \langle h_{0x}\rangle
    &=&\frac{h^{(+)}}{\sqrt{2}}\sin\theta\int_0^1\int_0^\pi\int_0^{2\pi}dr' d\zeta d\phi ~r'^2 \sin\zeta \no\\[6pt]
&&\hspace{3cm}\times\left(\tan^2\zeta\sin^2\phi- 1 \right)\left[\cos(\bar{\epsilon} r' \cos\zeta)-\frac{\sin(\bar{\epsilon} r'\cos\zeta)}{\bar{\epsilon} r'\cos\zeta}\right]\nonumber\\[6pt]
    &=&\frac{h^{(+)}}{\sqrt{2}}\sin\theta F(\bar{\epsilon}) \ ,
    \end{eqnarray}
where we defined
\begin{eqnarray}
{\bar F}(\bar{\epsilon})\equiv\frac{\pi\left(\bar{\epsilon} ^2-12\right){\rm Si}(\bar{\epsilon})}{4\bar{\epsilon}}+\frac{\pi\left(\bar{\epsilon} ^2-30\right)\cos (\bar{\epsilon})}{4\bar{\epsilon}^2}+\frac{\pi\left(\bar{\epsilon} ^2+30\right) \sin (\bar{\epsilon} )}{4\bar{\epsilon}^3}\,.
\end{eqnarray}
For $i=y$, we can deduce 
\begin{eqnarray}
    k_i h_{jk}^{\rm TT} x^jx^k&=&k_y h_{jk}^{\rm TT} x^jx^k=0\\[6pt]
    k_k h_{ji}^{\rm TT} x^jx^k
    &=&k\sin\theta\left(h_{xy}xx + h_{yy}yx+h_{zy}zx\right) + k\cos\theta\left(h_{xy}xz + h_{yy}yz+h_{zy}zz\right)\no\\[6pt]
&=&k\sin\theta\left(\frac{h^{(\times)}}{\sqrt{2}}\cos\theta xx - \frac{h^{(+)}}{\sqrt{2}}yx - \frac{h^{(\times)}}{\sqrt{2}}\sin\theta  zx\right)\no\\[6pt]
    &&+ k\cos\theta\left(\frac{h^{(\times)}}{\sqrt{2}}\cos\theta xz - \frac{h^{(+)}}{\sqrt{2}}yz - \frac{h^{(\times)}}{\sqrt{2}}\sin\theta  zz\right)\,.
    \end{eqnarray}
Again, the cross terms vanish after averaging. Hence, we obtain
    \begin{eqnarray}
    h_{0y}
    &=&\omega_gk\frac{h^{(\times)}}{\sqrt{2}}\cos\theta\sin\theta\left(x^2-z^2\right)\left[\frac{\cos(\bm{k}\cdot\bm{x})}{(\bm{k}\cdot\bm{x})^2}-\frac{\sin(\bm{k}\cdot\bm{x})}{(\bm{k}\cdot\bm{x})^3}\right]\no\\[6pt]
    &=&\frac{h^{(\times)}}{\sqrt{2}}\cos\theta\sin\theta\left(\tan^2\zeta\cos^2\phi- 1 \right)\left[\cos(kr\cos\zeta)-\frac{\sin(kr\cos\zeta)}{kr\cos\zeta}\right]\,.
    \end{eqnarray}
After the averaging, we have
    \begin{eqnarray}
    \langle h_{0y}\rangle&=&\frac{h^{(\times)}}{\sqrt{2}}\cos\theta\sin\theta {\bar F}(\bar{\epsilon})\,.
    \end{eqnarray}
Finally, we obtain
    \begin{eqnarray}
    \langle \Omega_g\rangle &=& \left\langle\bm{d}^{(34)}\cdot\bm{E}\right\rangle
    = {d^{(34)}}^i\varepsilon^{ijk}\langle h^0{}_j\rangle B^k\no\\[6pt]
    &=&d^{(34)}_{x}\langle h_{0y}\rangle B_z-d^{(34)}_{y}\langle h_{0x}\rangle B_z\no\\[6pt]
    &=&\frac{B_z}{\sqrt{2}}\sin\theta\left[d^{(34)}_x h^{(\times)}\cos\theta-d^{(34)}_{y}h^{(+)}\right]{\bar F}(\bar{\epsilon})\,.
    \end{eqnarray}
Note that ${\bar F}({\bar\epsilon})=F(\epsilon)$ and $\bar{\epsilon}=2\pi\epsilon$.
%==========================================================
%%%%%%%%%%%%%%%%%%%%%%%%%%%%%%%%%%%%%%%%%%%%%%%%%%%%%%%%%%%%%%%%%
\bibliography{Rydberg} 
\bibliographystyle{unsrt}
\end{document}